\documentstyle[twocolumn,aps,prd,epsf,epsfig]{revtex}

\begin{document}

\twocolumn[\hsize\textwidth\columnwidth\hsize
\csname@twocolumnfalse\endcsname
\draft
\title{Dynamics of laser-induced lattice formation in Bose-Einstein
condensates}
\author{M.~Kalinski$^1$, I.E.~Mazets$^{2,3}$, G.~Kurizki$^2$,
B.A.~Malomed$^4$, K.~Vogel$^1$ and W.P.~Schleich$^1$}
\address{{\setlength{\baselineskip}{18pt}
$^1$\,Abteilung f\"ur Quantenphysik, Universit\"at Ulm, 
Albert-Einstein-Allee 11, D-89069 Ulm, Germany,\\
$^2$\,Department of Chemical Physics, Weizmann Institute of Science,
Rehovot 76100, Israel,\\
$^3$\,Ioffe Physico-Technical
Institute, St.Petersburg 194021, Russia,\\
$^4$\,Department of Interdisciplinary Studies, Faculty of Engineering,
Tel-Aviv University, Tel-Aviv 69978, Israel
}}
\maketitle

\begin{abstract}
We present variational and numerical solutions of the generalized
Gross-Pitaevskii equation for an atomic Bose-Einstein condensate
with time-dependent, laser-induced dipole-dipole interactions (LIDDI).
The formation of a spatially modulated ground state due to LIDDI in
an elongated (cigar-shaped) BEC is investigated. We find that modulated
densities can emerge if the laser intensity exceeds a critical value,
which depends on the linear density of atoms in the BEC.
\\
\end{abstract}
\vskip1pc]

\section{Introduction}

The confinement of atomic Bose-Einstein condensates (BECs) in optical
traps \cite{o98,chap} and lattices \cite{ol} is 
commonly provided by a spatial
inhomogeneity of the laser intensity, which gives rise to gradient (or
dipole) forces acting on single atoms. Completely different dynamical
effects are generated in BECs by interatomic laser-induced dipole-dipole
interactions (LIDDI) \cite{sg,xp}. These effects take place even if the
atoms are illuminated by a plane electromagnetic wave. One such effect,
predicted by means of a 1D variational  ansatz, is that LIDDI caused by
an off-resonant circularly 
polarized plane wave traveling along a cigar-shaped
BEC results in a spatially periodic density modulation \cite{xp}.
In the present paper we discuss how such density modulations can emerge
from a uniform BEC when the laser intensity reaches sufficiently large
values.

\subsection{Supersolids versus BECs in optical lattices}

It is tempting to refer to such a density-modulated BEC as a supersolid.
Indeed, Ref.~\cite{xp} used this term to emphasize the presence of both
spatial modulation and phase coherence. Moreover, there exists also close
connection to a BEC in an optical lattice \cite{ol}. For this reason we
dedicate this subsection to a comparison of our density-modulated BEC to
supersolids and BECs in optical lattice.

Superfluid systems with spatially modulated densities have a long history.
The original question, however, was not whether a homogeneous BEC can be
spatially modulated.  Instead, the discussion started from the question
whether a solid composed of bosonic atoms can be superfluid. First, Penrose
and Onsager \cite{po56} demonstrated that Bose-Einstein condensation is
impossible in an ideal crystal. However, later it became clear \cite{vac,cheng}
that if there were vacancies present 
in a crystal then Bose-Einstein condensation
would be possible. The work of Chester \cite{ch70} and Leggett \cite{l70} led
to a deeper understanding of of Bose-condensed solids. However, the authors of
Refs.~\cite{po56,vac,cheng,ch70,l70} concentrated on crystals where the
occupation number of a lattice site never exceeds 1. Thus the estimation for
the ratio between the superfluid density and the normal density in such a
system was rather pessimistic ($\sim 10^{-4}$) \cite{l70}. In contrast,
the situation for liquid $^4$He is more favorable. For example,
Pitaevskii \cite{p84} predicted the formation of a standing wave in helium
(below the critical temperature) flowing at a supercritical velocity in a
capillary tube. The systems studied in the references mentioned above are
quite different from a Bose-condensed dilute gas.

However, the studies of Pomeau and Rica \cite{pom} and G\'{o}ral
et.\ al \cite{goral} are closer to our work. Pomeau and Rica analyzed what
could happen if the roton minimum in the liquid helium spectrum reaches the
horizontal axis and even if the homogeneous system would become unstable at
the corresponding wavelength. Their finding is that as soon as the uniform
ground state becomes unstable, another ground state, 
spatially modulated, comes
to existence, instead. Pomeau and Rica call such a density-modulated state
``supersolid'', an expression already used by Cheng \cite{cheng}.

G\'{o}ral et.\ al discussed quantum phases of dipolar bosons in an optical
lattice. Their system consists of a dilute atomic gas where the atoms interact
not only via the short-range $s$-wave scattering, but also -- due to their
large magnetic dipole moments -- via long-range dipole-dipole interaction.
This dipole-dipole interaction is somewhat similar to the LIDDI, except that 
the magnetic dipole-dipole interaction is purely static, varies as 
the inverse distance cubed, and does not exhibit oscillatory 
(retardation) features, which are esssential in considerations of the 
LIDDI \cite{sg,xp}. They found that in an optical lattice such a system
can have various phases, including a supersolid phase. We conclude our brief
discussion of supersolids by mentioning that recently a supersolid phase of
$^4$He may have been observed \cite{kim}.

Another superfluid system that displays spatially modulated densities is a
BEC in an optical lattice. Here up to several hundred atoms can occupy one
lattice site. As first demonstrated by Anderson and Kasevich \cite{and99},
macroscopic quantum interference can be observed in such systems. As we show
in the present paper, in our system many atoms contribute to a maximum in
the density. Therefore, we expect our system to have similar coherence
properties as a BEC in an optical lattice.

Due to its analogy to BECs in optical lattices we may call the effect
considered in the present paper a spontaneous formation of an optical lattice
in a laser-illuminated BEC. Such a lattice is self-sustained, since it is
formed by interfering incident and coherently backward scattered light waves.

\subsection{Formulation of the problem}

While Ref.~\cite{xp} has suggested the intriguing possibility of
``spontaneous laser-induced periodicity'' in a BEC, it has left open
basic questions related to their observability:
(a) Is the suggested 1D variational steady-state solution suitable
for a cigar-shaped BEC, or should we expect radial density modulation
as well?
(b) Is a spatially modulated ground state -- the signature of spontaneous
laser-induced periodicity -- obtainable from a uniform initial ground state
by gradually increasing the laser intensity, even though it must compete with
evaporation via Rayleigh scattering and possible non-adiabatic effects?
(c) Is the this density modulation distinguishable from that obtained by
the superradiant response of a BEC \cite{sr}?

In this work we aim at a realistic investigation of laser-induced formation
of spatially modulated densities in a BEC that would provide clear answers
to the questions raised above. This is accomplished through an exhaustive
numerical investigation that fully accounts, for the first time, for the 3D
geometry of the cigar-shaped sample and the time-dependent dynamics of the
formation process. Our numerical results show that in the proposed setting
spontaneous laser-induced periodicity is experimentally feasible. They are
supplemented by a simple variational approximation, which is shown to explain
the essential features of the numerical results and their stability. We also
compare and contrast the differences of the present LIDDI effects to
superradiant light scattering by a BEC \cite{sr}.

\subsection{Outline of the paper}

Our paper is organized as follows: In Sec.~\ref{sec:GPE} we introduce
a Gross-Pitaevskii equation which takes into account the laser-induced
dipole-dipole interaction and discuss the interplay between the long-range
dipole-dipole interaction and the short-range $s$-wave scattering.
Section~\ref{sec:VAR} is devoted to a variational approach to the problem.
We assume a cigar-shaped BEC with a tight radial confinement and arrive at
a one-dimensional equation. We use variational calculations to obtain
approximate stationary solutions of this equation. In Sec.~\ref{sec:NUM}
we solve the full Gross-Pitaevskii equation numerically and find time-dependent
as well as stationary solutions. Based on these results we discuss a possible
experiment in Sec.~\ref{sec:DIS}. Furthermore, we comment on the depletion
of the condensate in such an experiment and mention other strategies to
obtain spatially modulated densities. We also compare density-modulated BECs
formed by LIDDI to coherent atom recoil lasing. We conclude by summarizing
our results in Sec.~\ref{sec:CON}.

\section{Gross-Pitaevskii equation}
\label{sec:GPE}

The time evolution of the BEC order parameter $\Psi $ is governed by the 3D
Gross-Pitaevskii equation
\begin{eqnarray}
i\hbar \frac{\partial }{\partial t}\Psi ({\bf r},t) &=&-\frac{\hbar ^{2}}{2m}
\nabla ^{2}\Psi ({\bf r},t)+V_{{\rm trap}}({\bf r})\Psi ({\bf r},t)
\nonumber \\
&&+\frac{4\pi \hbar ^{2}a_{s}}{m}\left| \Psi ({\bf r},t)\right| ^{2}\Psi
({\bf r},t)  \nonumber \\
&&+\int d^{3}{\bf r}^{\prime }\,\left| \Psi ({\bf r}^{\prime },t)\right|
^{2}V_{dd}({\bf r}-{\bf r}^{\prime })\Psi ({\bf r},t).  \label{eq1}
\end{eqnarray}
Here
\begin{equation}
V_{{\rm trap}} \approx \frac{m}{2}\,\left[
\omega_{\perp }^{2} (x^{2}+y^{2})+\omega _{\Vert}^{2}z^{2}
\right]
\end{equation}
denotes the external trapping potential and $m$ and $a_{s}$ are the atomic
mass and the $s$-wave scattering length due to short-range interatomic forces,
respectively. The last term in Eq.~(\ref{eq1}), which comes in addition to the
terms of the usual Gross-Pitaevskii equation \cite{gp,hp}, accounts for the
long-range LIDDI. Note that $\Psi ({\bf r},t)$ is normalized to the number
of atoms.

For a circularly-polarized laser wave propagating along the $z$ axis, the
kernel of this effective interaction reads \cite{xp}
\begin{eqnarray}
V_{dd}({\bf r}) &=&U_{0}\left[ \left( 3\cos ^{2}\vartheta -1\right) \frac{
\cos qr+qr\sin qr}{(qr)^{3}}\right.  \nonumber \\
&&\left. -\left( \cos ^{2}\vartheta +1\right) \frac{\cos qr}{qr}\right] \cos
(qr\cos \vartheta ).  \label{vdd}
\end{eqnarray}
Here, the direction of the interatomic-separation vector ${\bf r}$ relative
to the $z$ axis is determined by the polar and azimuthal angles $\vartheta$
and $\varphi$, respectively, and $q$ denotes the laser wave number. The
characteristic energy $U_{0}$ corresponding to the interaction of two atoms
separated by $qr\sim 1$ is proportional to the laser intensity. On the other
hand, the rate $\Gamma _{Ray}$ of the incoherent (Rayleigh) photon scattering
per atom is also proportional to the laser intensity which provides us with
the connection $U_0 = \frac{3}{4}\hbar \Gamma _{Ray}$.

From the outset, we must address two questions:
(i) Can we use in Eq.~(\ref{eq1}) the {\em bare} potential $V_{dd}({\bf r})$
of Eq.~(\ref{eq1})?
(ii) To what extent does the interplay between the long-range potential
$V_{dd}({\bf r})$ and the short-range interatomic potential change the
value of $a_{s}$ in Eq.(\ref{eq1})?

To answer the first question we recall that according to Belyaev's theory of
effective interactions \cite{be}, the bare potential should be replaced by an
effective potential, such that the scattering amplitude produced by the
effective potential {\em in the first Born approximation} coincides with the
{\em exact} scattering amplitude generated by the bare potential. We have
performed such a calculation for the isotropic part $V_i(r)$ of the LIDDI
potential, which has the asymptotic forms $V_i(r)\approx -U_{0}(qr)^{-2}\sin (2qr)$
for $qr\gg 1$, and $V_i(r) \approx -\left( 22/15\right) U_{0}(qr)^{-1}$ for
$qr\ll 1$ \cite{iso}. We have found, that, for this isotropic part, the Born
scattering amplitude is always close to the exact one. We conjecture that the
same is true for the entire (anisotropic) potential $V_{dd}({\bf r})$.

To address the second question we first recall \cite{yy} that the part of
$V_{dd}$ being proportional to $(3\cos ^{2}\vartheta -1)r^{-3}$ has no
significant influence on $a_{s}$ as long as we are far from the shape
resonance. In contrast, Ref.\ \cite{lnd} shows that the addition of the
Coulomb interaction to a short-range potential, may dramatically change the
$s$-wave scattering properties. Since for short distances the isotropic part
of $V_{dd}$ has Coulomb character, this contribution may have some impact on
the $s$-wave scattering length.

To clarify this issue, we have numerically calculated the phase of the
scattered wave, combining the isotropic part of $V_{dd}$ and the short-range
potential. The latter has been taken into account by the boundary condition
$\chi (r)^{\prime }/\chi(r) |_{r\rightarrow 0}=-1/a_{s}$, for the solution
$\chi(r)$ of the radial Schr\"{o}dinger equation describing the scattering
of two atoms. We have not found any dramatic change of $a_{s}$ throughout the
relevant range of relative momenta of the colliding atoms. This result is due
to the fact that the gravity-like part of $V_{dd}({\bf r})$ has a cutoff at
$r\sim q^{-1}$, which is much shorter than the corresponding Bohr's
radius \cite{sg} for experimentally feasible laser intensities. Thus, we may
use, with reasonable accuracy, the same value of $a_{s}$ as without the laser
field.

\section{Variational calculations}
\label{sec:VAR}

Before we solve Eq.~(\ref{eq1}) numerically we first want to gain a
qualitative understanding of the dynamics of the system described by
Eq.~(\ref{eq1}). For this purpose we have developed a quasi-1D variational
approximation which we discuss in this section. This approximation already
predicts spontaneous laser-induced periodicity, that is, a spatially modulated
ground state, when the laser intensity exceeds a threshold value.

\subsection{Ansatz}

We start from the fact that Eq.~(\ref{eq1}) can be derived
from the Lagrangian $L=\int d^{3}{\bf r}\,{\cal L}$ with
\begin{eqnarray}
{\cal L} &=&\frac{i\hbar }{2}\left[ \Psi^{\ast} ({\bf r},t)\frac{\partial }
{\partial t}\Psi({\bf r},t)-\Psi({\bf r},t)\frac{\partial}
{\partial t}\Psi^{\ast} ({\bf r},t)\right]   \nonumber \\
&&-\frac{\hbar ^{2}}{2m}[\nabla \Psi ({\bf r},t)][\nabla \Psi ^{\ast }({\bf r}
,t)] -V_{{\rm trap}}({\bf r})|\Psi ({\bf r},t)| ^2
\nonumber \\
&& -\frac{2\pi \hbar ^{2}a_{s}}{m}\left| \Psi ({\bf r},t)\right| ^{4}
\nonumber \\
&& -\frac{1}{2}\int d^{3}{\bf r}^{\prime }\left| \Psi ({\bf r}^{\prime
},t)\right| ^{2}V_{dd}({\bf r}-{\bf r}^{\prime })\left| \Psi ({\bf r}
,t)\right| ^{2}.  \label{eq2}
\end{eqnarray}
We assume a cigar-shaped condensate where the radial confinement is so
tight that, regardless of the atomic density, only the ground state of
the transverse motion is occupied. This assumption allows us to approximate
\cite{Salasnich} the transverse distribution of the order parameter by
\begin{equation}
\Psi ({\bf r},t) \approx (\sqrt{\pi }\sigma)^{-1}
\exp [-(x^{2}+y^{2})/(2\sigma^{2})]\psi (z,t)
\label{ap3}
\end{equation}
with $\sigma \equiv \sqrt{\hbar /(m\omega _{\perp })}$.
This approximation reduces the full Gross-Pitaevskii equation to
a one-dimensional equation for the longitudinal part $\psi(z,t)$.

Since for
\begin{equation}
|z|\gg q^{-1}\sim \sigma\,\,_{\sim }^{>}\,\,\sqrt{x^{2}+y^{2}}
\end{equation}
the dipole-dipole potential takes the form
\begin{equation}
V_{dd}({\bf r})\approx -2U_{0}(qz)^{-1}\cos ^{2}\left( qz\right)
\end{equation}
we expect a spatial structure with the period $\pi /q$ to appear in the ground
state. Furthermore, we assume for our variational approximation that the
trapping potential $V_{\rm trap}({\bf r})$ does not confine the BEC in
$z$-direction, that is we assume $\omega_{||} \approx 0$. We therefore pursue
the 1D variational {\em ansatz}
\begin{equation}
\psi (z,t)=a(t)\cos \left( 2qz\right) +c(t),  \label{ap1}
\end{equation}
where $a(t)$ and $c(t)$ are complex variables.

To stabilize the BEC against long-wavelength perturbations \cite{xp}, we
must also require that the short-range interatomic repulsion, characterized
by a positive scattering length, $a_{s}>0$, is stronger than the net
laser-induced attraction of two atoms separated by $r$ $_{\sim }^{<}\,\,q^{-1}$.
Moreover, we argue that this short-range repulsion term does not affect the
dynamics at the wavelength $\sim \pi /q$, since, for moderate BEC densities,
$\pi /q$ is smaller than the BEC healing length, which is determined by the
BEC density and $a_{s}$. Only under these conditions the variational ansatz
Eq.~(\ref{ap1}) describes the dynamics of the condensate appropriately.

\subsection{Equations of motion}

We now derive equations of motion for the coefficients $a(t)$ and $c(t)$.
For this purpose we substitute Eq.~(\ref{ap3}) with Eq.~(\ref{ap1}) into
the Lagrangian density  Eq.(\ref{eq2}) and perform the integration. Here we
consider a ``cigar'' of size $\ell $ in the $z$-direction and neglect
interaction terms that do not contain the large logarithmic factor
$\Lambda \approx 2\log [\ell /(2\sigma)]$, which diverges as
$\ell \rightarrow \infty $. The effective Lagrangian density
\begin{eqnarray}
\frac{2L}{\ell \hbar } &=&i(c^{\ast}\dot{c}-\dot{c}^{\ast}c)
+\frac{i}{2} (a^{\ast}\dot{a}-\dot{a}^{\ast}a)-\Omega_{q}a^{\ast}a
\nonumber \\
&&+g\left[ (c^{\ast}c+\frac{1}{2}a^{\ast}a)^{2}
+\frac{1}{4}(c^{\ast }a+a^{\ast }c)^{2}\right] ,
\label{LLL}
\end{eqnarray}
contains the recoil frequency $\Omega _{q} \equiv 2\hbar q^{2}/m$ associated
with the momentum transfer $2\hbar q$, and $g \equiv (\hbar q)^{-1}U_{0}\Lambda$
is the effective dipole-dipole coupling constant. The Euler-Lagrange equations
corresponding to Eq.~(\ref{LLL}) read
\begin{eqnarray}
\dot{a} &=&-i\Omega_{q}a + i\left( g/2\right)
(3|c|^{2}a+|a|^{2}a+c^{2}a^{\ast }),  \label{s1a} \\
\dot{c} &=&i\left( g/4\right) (4|c|^{2}c+3|a|^{2}c+a^{2}c^{\ast })
\label{s1b}
\end{eqnarray}
and thus the effective Hamiltonian density
\begin{equation}
\frac{2 H}{\ell \hbar}
=\Omega _{q}|a|^{2}-\frac{1}{4}g\left[ |a|^{4}+6\left| ca\right|
^{2}+4|c|^{4}+\left( ac^{\ast }\right) ^{2}+\left( a^{\ast }c\right) ^{2}
\right] \label{H}
\end{equation}
is a constant of the motion. Another dynamical invariant of Eqs.~(\ref{s1a})
and (\ref{s1b}) is the linear density of atoms
\begin{equation}
{\cal N} \equiv |c|^{2}+ |a|^{2}/2\,.
\end{equation}
Hence, we are left with three independent real variables, which can be chosen
as
\begin{eqnarray}
n & \equiv & |a|^{2}-2\,|c|^{2},
\\
s & \equiv & 2\sqrt{ 2}\,{\rm Re}\,(a^{\ast }c),
\\
p & \equiv & -2\sqrt{2}\,{\rm Im}(a^{\ast }c).
\end{eqnarray}
We then arrive at the set of equations
\begin{eqnarray}
\dot{n} & = & \left( g/2\right) sp,
\label{s2c:1}
\\[1.5ex]
\dot{s} & = & \Omega _{q}p,
\label{s2c:2}
\\
\dot{p} & = & -s \left(\Omega _{q} + \frac{1}{2} g n \right)
\label{s2c:3}
\end{eqnarray}
which is similar to those governing the evolution of the Bloch
vector \cite{blo}. In particular, $s$ is the analog of the Rabi
frequency. A general solution of these equations can be found
in terms of elliptic functions.

\subsection{Stationary solutions}

We now focus on stationary solutions, that is fix points of
Eqs.~(\ref{s2c:1})-(\ref{s2c:3}) and discuss their stability. We recognize
that the choice $s=p=0$ is the simplest stationary solution. The definitions
of $s$ and $p$ imply that either $a$ or $c$ are zero. In case $a=0$ and
$c \ne 0$ we find from Eq.~(\ref{ap1}) a spatially uniform solution. From
the definition of $n$ and the dynamical invariant ${\cal N}$ we arrive at
$n_{u}=-2{\cal N}$. The case $c=0$ and $a \ne 0$ corresponds to a purely
sinusoidal $\psi (z)$ leading to $n_{s}=2{\cal N}$. These two fix points
exist for all values of the parameters.

The third fix point results from the choice $p=0$ and $s \ne 0$. This
condition implies $\Omega_{q} + g n /2 = 0$, which leads to
$n_{ss}=-2\Omega_{q}/g$. Moreover, $p=0$ and $s \ne 0$ demands $a \ne 0$
and $c \ne 0$ and therefore corresponds to a spatial modulation on top of
a constant background. From the definitions of $s$ and $p$ we find
\begin{equation}
s_{ss}=2 \sqrt{{\cal N}^{2} -(\Omega_{q}/g)^{2}} \,\,.
\label{co1}\end{equation}
Since $s$ has to be real, this fix point exists provided
$g{\cal N}>\Omega_{q}$, that is the product of $g$ determined by the laser
intensity and the atom density ${\cal N}$ exceeds a threshold value,
determined by the recoil frequency $\Omega_{q}$.

We proceed with a standard stability analysis and find the perturbation
eigenfrequencies
\begin{eqnarray}
\eta_{u} & = & \sqrt{\Omega_{q}(\Omega _{q}-g{\cal N})}\,\,,
\\
\eta_{s} & = & \sqrt{\Omega_{q}(\Omega _{q}+g{\cal N}/2)}\,\,,
\\
\eta_{ss} & = &\sqrt{(g{\cal N})^{2}-\Omega _{q}^{2}}\,\,.
\end{eqnarray}
From the expressions for $\eta_{ss}$ and $\eta_{u}$ we recognize that the
spatially modulated solution is stable throughout its range of existence,
whereas the uniform solution is unstable in the same region. The uniform
solution is stable if $g{\cal N}<\Omega_{q}$, that is when the modulated
solution does not exist. The sinusoidal solution is always stable. However,
its energy determined by the Hamiltonian (\ref{H}) is larger than the one
corresponding to the other stable solution. Hence, the modulated solution
represents the {\em spatially modulated ground state} of the quasi-1D BEC
illuminated by laser light. If the condition $g{\cal N}>\Omega_{q}$ is not
met, then the ground state corresponds to the uniform solution. The latter
becomes unstable with the increase of the laser intensity and/or the BEC
density beyond the bifurcation point $g{\cal N}=\Omega_{q}$.

\section{Numerical results}
\label{sec:NUM}

We now turn to the numerical solution of Eq.~(\ref{eq1}), using the
second-order split-operator method. The integral convolution of the
kernel $V_{dd}$ with the matter density $|\Psi|^2$ in the last term
of Eq.~(\ref{eq1}) is calculated using the Fast Fourier Transform (FFT)
technique \cite{ij2}.

\begin{figure}
\begin{center}
\centerline{\psfig{file=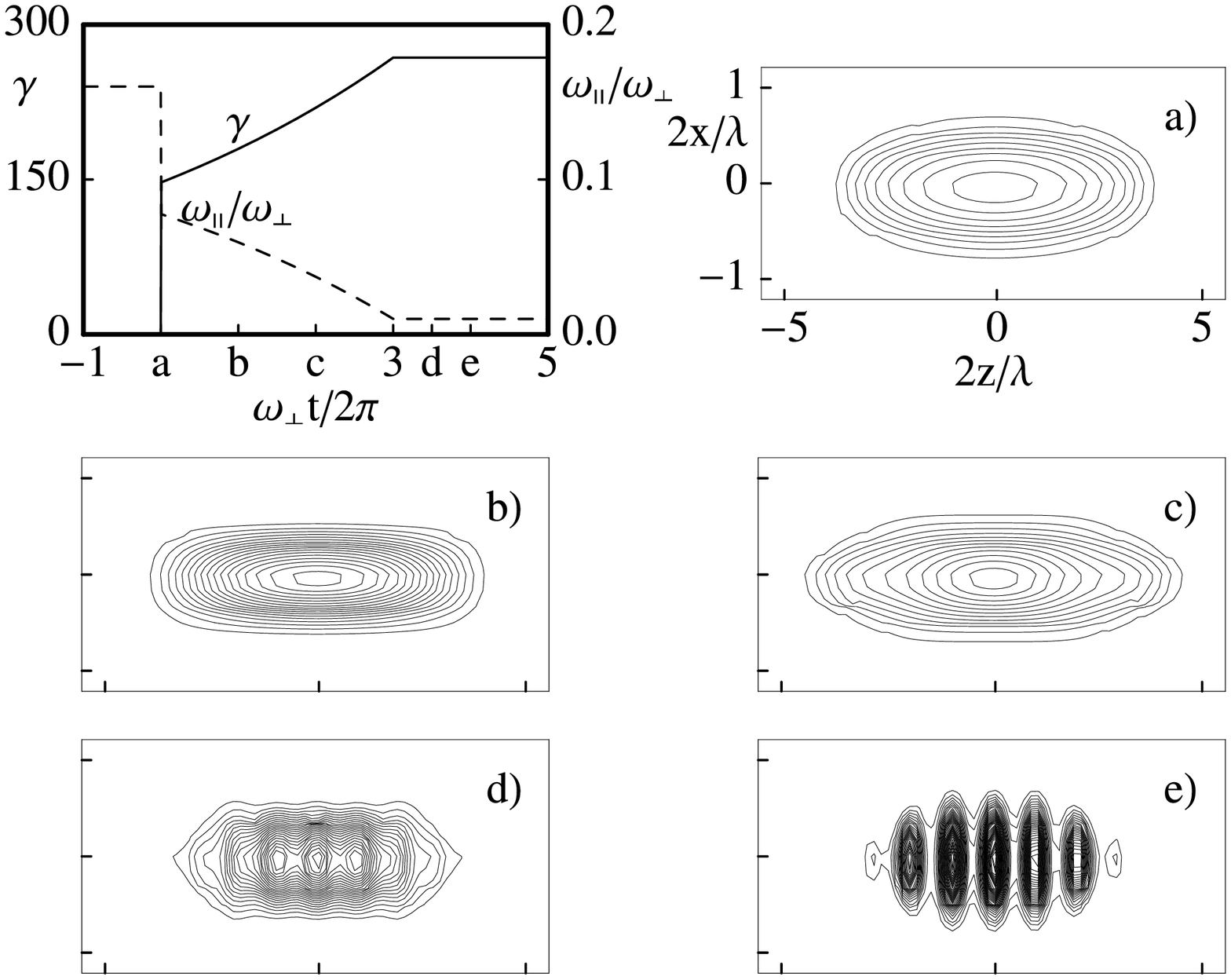,width=7cm}}
\end{center}
\begin{caption}
{
Spontaneous laser-induced periodicity due to LIDDI. A BEC is illuminated
by a circularly polarized laser beam oriented along the $z$-axis. The
upper left plot shows the time dependence of the laser intensity and the
longitudinal trap frequency. The time evolution of the density is shown
in a) -- e) for various times marked in the upper left plot. We have used
scaled quantities where time is measured in periods of the transverse motion
and lengths are measured in units of half a laser wavelength. The dimensionless
dipole-dipole interaction strength $\gamma$ is defined by
$\gamma=2\pi^2 N U_0/(\hbar \Omega_q)$. The other parameters are
$\pi^2 \omega_{\perp}/\Omega_q = 5$ and $N q a_s = 25$. Here $N$ is the
number of particles and $\Omega_q$ denotes the recoil frequency introduced
in Eq.\,(\ref{LLL}). Note that we have used different scalings for the radial
and the longitudinal direction.
}
\label{fig:time}
\end{caption}
\end{figure}

To circumvent the difficulties with the singularity of $V_{dd}$ at $r=0$
we introduce a small cutoff radius $r_c\sim 10^{-3}q^{-1}$. We calculate
the time evolution of the condensate starting from the ground state of
the BEC in an elongated ($\omega _{\perp }\gg \omega_{\Vert }$) trap and
switch on the laser intensity. In order to maximize the overlap of the
initial (cigar-shaped) and final (spatially modulated) ground state, we
try to keep the size of the condensate fixed when we increase the intensity
of the laser field. In our numerical calculations we change the trapping
potential by decreasing $\omega_{\Vert}$, so as to compensate for the
laser-induced compression (electrostriction) of the condensate, see upper
left plot in Fig.\,\ref{fig:time}. Another way is to ramp-up the scattering
length by approaching a Fesh\-bach resonance at a rate similar to the laser
switch-on.

Figure \ref{fig:time} clearly shows spontaneous laser-induced periodicity:
After we have increased the laser intensity and decreased the longitudinal
frequency sufficiently enough, a spatially modulated density emerges.

\begin{figure}
\begin{center}
\centerline{\psfig{file=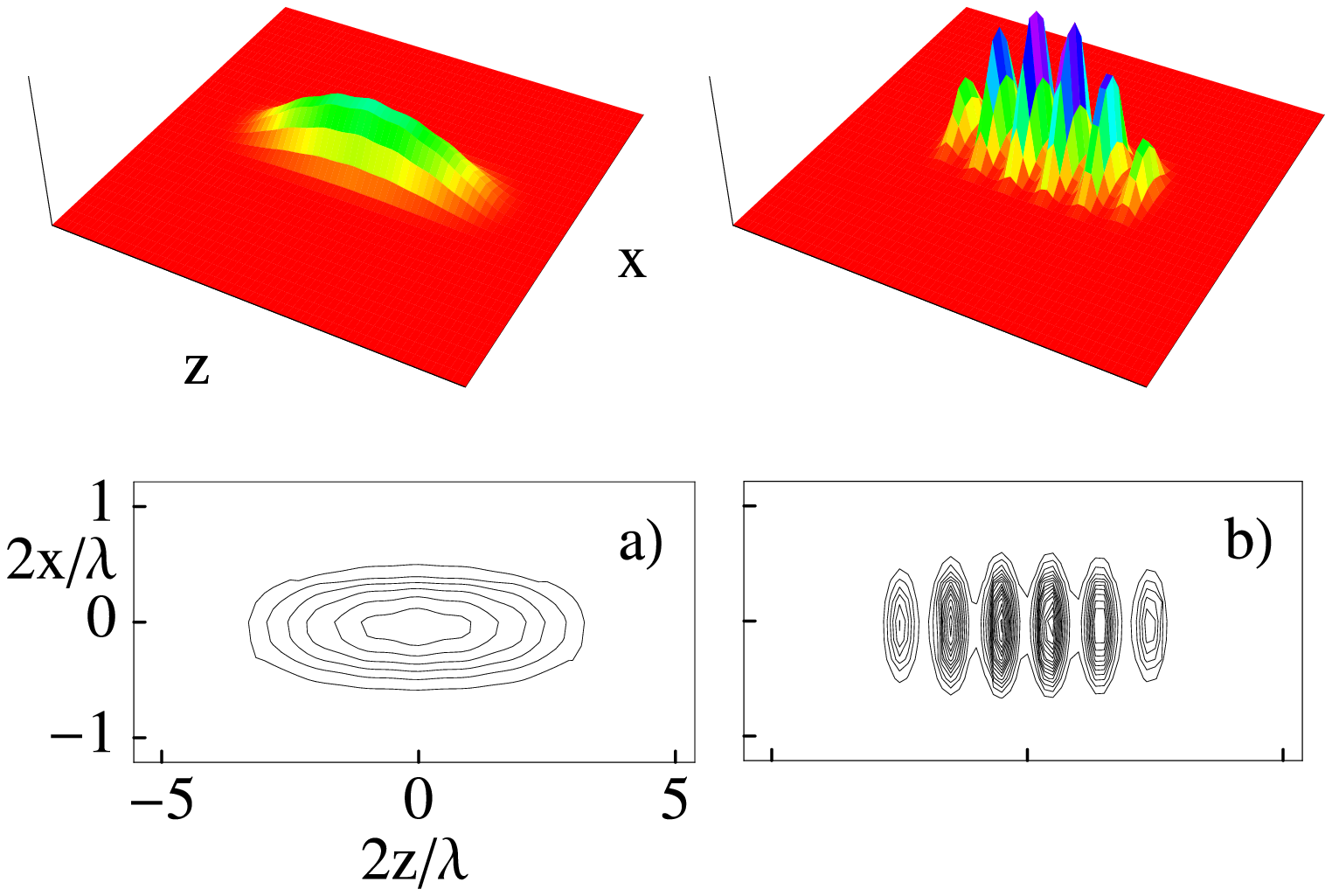,width=7cm}}
\end{center}
\begin{caption}
{Stationary solutions of Eq.~(\ref{eq1}) for a scaled laser intensity of
$\gamma=234.55$ (a) and $\gamma=268.06$ (b). The other parameters are the
final values of Fig.\,\ref{fig:time}. Only if the laser intensity is large
enough the density is spatially modulated. Again, we have used different
scalings for the radial and the longitudinal direction.
}
\label{fig:steady}
\end{caption}
\end{figure}

Moreover, we perform also imaginary time calculations to find the
quasi-stationary density distribution for two different values of the laser
intensity. Our numerical results show that these density distributions are
spatially modulated only if the laser intensity exceeds a certain critical
threshold, see Fig.\,\ref{fig:steady}. For the larger laser intensity we find
a quasi-stationary distribution which has a similar shape as the final state
in Fig.\,\ref{fig:time}e. It is noteworthy that {\em no oscillatory structure
is numerically observed in the radial direction}.

We find qualitative agreement between the variational approach and the
numerical results. Our variational approach predicts spontaneous laser-induced
periodicity for $g{\cal N}/\Omega_q > 1$, whereas from our numerical data we
find a transition to a spatially modulated density at $g{\cal N}/\Omega_q
\approx 2.5$. The agreement is not too bad considering the fact that there
are only few density peaks in the numerically obtained density distribution.

For numerical reasons we had to limit the length of the cigar-shapedBEC
to few laser wavelengths. We have not found any spatially modulated density,
neither in the longitudinal nor in the transverse direction, if the
condensate is significantly wider than approximately half a wavelength.
Our numerical results therefore do not allow us to answer the question
whether we have to confine the radial direction to approximately half a
laser wavelength in order to observe spontaneous laser-induced periodicity,
or whether a sufficiently large aspect ratio can already lead to spontaneous
laser-induced periodicity.

\section{Discussion}
\label{sec:DIS}

A possible experimental realization could involve BECs with a linear density
of approximately 630 atoms/$\mu$m. For $^{87}$Rb the laser detuning may be
chosen to be $2\pi \times 6.5$ GHz, the resonant wavelength being
0.795~$\mu $m. A pronounced spatially modulated density should then appear at
the laser intensity $I_{c}=38$~mW/cm$^2$, that corresponds to
$\Gamma _{Ray}=133$~s$^{-1}$. Our variational approach [Eq.(\ref{co1})] gives
$I_c\Lambda =34$~mW/cm$^2$, where $\Lambda \approx 2\log [\ell /(2\sigma)]$.

We may assume exponential growth of the laser intensity at the rate of
$\sim 10^3$~s$^{-1}$ and simultaneous decrease of the longitudinal frequency
$\omega_{\Vert}$.  Under these conditions, the collective oscillations in the
BEC, which are excited due to non-adiabaticity of switching of the intensity,
are found to be small. The total depletion of the BEC due to incoherent photon
scattering is around 50 percent. Thus one can see that spontaneous laser-induced
periodicity is possible, although accompanied by considerable losses of the
number of condensed atoms.

\subsection{Quantum depletion}

We are aware of the problem of the quantum depletion of the condensate.
Indeed, in a recent paper \cite{qd1} the quantum depletion has been
calculated in the Popov approximation for a case of a laser-radiated BEC
in a geometry alternative to that of our present work. In contrast to
the present geometry, where a circularly polarized plane wave is traveling
along a cigar-shaped BEC, Ref.~\cite{qd1} considers a situation in which
the wave vector and the polarization vector of a linearly polarized laser
beam are perpendicular and parallel to the major axis of a cigar-shaped BEC,
respectively.  As the laser-induced roton-like minimum in the BEC elementary
excitation spectrum tends to touch the horizontal axis, the quantum depletion
growth rapidly, the peak being centered at the roton-minimum position in
the momentum space. However, in the present paper we consider, unlike
Ref.~\cite{qd1}, not a stationary but a time-dependent regime. The time of
formation of the distribution of the above-condensate atoms in a system
where both the short-range ($s$-wave scattering) and long-range (LIDDI)
interactions are present is an important question on its own. However, we
assume that the increase of the laser intensity in our case is rapid enough
to prevent the total depletion of the condensate before the spontaneous
laser-induced periodicity, that is, a transition from a uniform BEC to a
spatially modulated BEC, can be observed.

Nevertheless, a certain depletion is unavoidable. Thus, as soon as we
exceed the intensity threshold, the above-condensate atoms act as a seed
perturbation, which will grow exponentially at the wave vectors where the
spatially uniform solution becomes unstable. Therefore, the 
instability growth is concurrent to the emergence of the coherent spatial 
structure. Although the magnitude of
the seed perturbation has to be calculated beyond the
mean-field (Gross-Pitaevskii equation) approach, the
instability increment can be derived from the linearized Gross-Pitaevskii
equation \cite{hp}. In numerical simulations of unstable BECs the role of
seed perturbation is played by numerical noise \cite{qd2}. Stable numerical
algorithms avoid amplification of numerical rounding errors provided that the
numerical solution converges to the exact one, which is stable. However, if one
reaches an instability range, the numerical errors begin to grow due to the
properties of the equation being solved, and not due an unstable numerical
algorithm. Since our numerical solutions (see Fig.~1) display a stable spatial
structure, we may conclude that our transition from one stable branch (the
uniform one) to another (the spatially modulated one) is fast enough to
prevent an appreciable condensate depletion and the development of a
dynamical instability.

\subsection{Alternative approach}

A better strategy to obtain similar spatially modulated densities is to
first prepare the density-modulated ground state in an optical lattice
formed by a {\em far-detuned standing wave}. We then switch off the optical
lattice and simultaneously turn on a {\em running} wave which creates the
LIDDI. It is now this running wave that keeps the BEC in the initially
prepared density-modulated ground state. Although it is technically simpler
to create a spatially modulated self-sustained structure in a BEC starting
from a BEC confined in an external optical lattice, the way studied in the
present paper has its advantage. It clearly shows 
the {\em spontaneous} emergence
of spatially modulated structures when the laser intensity is gradually
increased above a threshold value.

\subsection{Laser-induced lattice formation versus coherent
atom recoil lasing}

We emphasize the key difference between the formation of spatially modulated
densities due to LIDDI in a laser-illuminated BEC and superradiant light
scattering by a BEC: The latter, related to coherent atom recoil lasing
(CARL) \cite{sr}, is essentially non-stationary, resulting in the generation
of phonons in a BEC. In contrast, the density modulations discussed in the
present paper are  quasi-stationary. Accordingly, under the conditions of
slow switch-on of the laser, we may expect a spatially-oscillatory ground
state without triggering superradiance. On the other hand, CARL has no
threshold in terms of the laser-beam intensity, unlike the effects discussed
above. Hence, a weak, abruptly switched-on laser pulse would yield {\em only}
superradiance.

\section{Conclusions}
\label{sec:CON}

To conclude, our 3D numerical calculations have substantiated the simple 1D
variational ansatz, predicting the gradual formation of an axial oscillatory
ground-state density modulation in a laser-illuminated cigar-shaped
BEC. Its origin can be unambiguously traced to LIDDI. Although the depletion
of the condensate due to incoherent photon scattering and due to instabilities
at the transition from a uniform condensate to a spatially modulated
condensate is considerable, we believe that the creation of spatial
modulations in a BEC is, as discussed in the present paper, experimentally
feasible. We emphasize the similarities between the spatial modulations due
to LIDDI and the modulations of a BEC in an optical lattice. We may call the
formation of a spatially modulated density a spontaneous formation of a
self-sustained optical lattice which is formed by interfering incident
and coherently backward scattered light waves.

\acknowledgments

The support of the German-Israeli Foundation, the EC (the QUACS RTN) and
Minerva is acknowledged. I.E.M. also thanks the Russian foundations
(RFBR 02--02--17686, UR.01.01.040, E02--3.2--287).


\begin{references}
\bibitem{iso}
The potential in this limit accounts for the ``artificial gravity'' that
may give rise to spherically-symmetric self-bound BEC states \cite{sg}.
\end{references}
\end{document}